# How do Data Journalists Design Maps to Tell Stories?

Arlindo Gomes, Emilly Brito, Luiz A. Morais and Nivan Ferreira

**Abstract**— Maps are essential to news media as they provide a familiar way to convey spatial context and present engaging narratives. However, the design of journalistic maps may be challenging, as editorial teams need to balance multiple aspects, such as aesthetics, the audience's expected data literacy, tight publication deadlines, and the team's technical skills. Data journalists often come from multiple areas and lack a cartography, data visualization, and data science background, limiting their competence in creating maps. While previous studies have examined spatial visualizations in data stories, this research seeks to gain a deeper understanding of the map design process employed by news outlets. To achieve this, we strive to answer two specific research questions: *what is the design space of journalistic maps?* and *how do editorial teams produce journalistic map articles?* To answer the first one, we collected and analyzed a large corpus of 462 journalistic maps used in news articles from five major news outlets published over three months. As a result, we created a design space comprised of eight dimensions that involved both properties describing the articles' aspects and the visual/interactive features of maps. We approach the second research question via semi-structured interviews with four data journalists who create data-driven articles daily. Through these interviews, we identified the most common design rationales made by editorial teams and potential gaps in current practices. We also collected the practitioners' feedback on our design space to externally validate it. With these results, we aim to provide researchers and journalists with empirical data to design and study journalistic maps.

**Index Terms**—design space, maps, news, visualization design, data visualization, data journalism

✦

## 1 INTRODUCTION

The *where* of a story is central in journalism [53]. Unsurprisingly, maps have long been a keystone of news articles [55]. They provide spatial context to locations or quantitative data related to specific subjects, including politics, economics, and climate, to name a few. Today, maps are commonplace in print, on the Web, and in TV news. Outlets have been applying them to inform and engage their audience, distinguishing themselves from competitors. As defined by Monmonier [38], *journalistic maps* are cartographic representations designed by journalists to visually support news narratives. Their design and implementation fall within Geojournalism [16], a subfield of data journalism [54] that integrates design and geographic information technologies to enhance news reporting and comprehension.

The production of journalistic maps presents several challenges. Journalists must deal with a trade-off between creating an engaging narrative while adapting the map design to audience's data literacy level, ensuring the content remains accessible for the general public. In large newsrooms with high article turnover, journalists also face intense time pressure to create maps, limiting opportunities for research and innovation in their design. On top of that, as journalistic cartography was developed independently from traditional cartography [40], most data journalists are self-taught [7] and do not have formal education on map design, data analysis, data visualization, or cartography to build journalistic maps. This can result in limited or ineffective use of possible map designs, which can cause distraction from the main narrative or even contribute to misinformation [11]. In addition, poorly designed maps can undermine audience trust and, consequently, damage the outlet's credibility [26]. Overall, these challenges highlight the need for a design framework that can both guide data journalists in creating effective map designs and inform visualization researchers about common practices and recurring problems in the field. While previous works have explored the structure and use of visualization in data journalism [23, 43], to the best of our knowledge, no prior work has focused on providing a comprehensive understanding of journalistic maps and the interconnected challenges involved in their design.

This study aims to help bridge this gap by proposing a design space for journalistic maps and providing an in-depth understanding of the design process of maps used in media outlets. To do this, we investigate two specific research questions: **(RQ1)** *What is the design space of journalistic maps?* and **(RQ2)** *How do editorial teams produce journalistic maps for news articles?* These questions were proposed to enable a comprehensive study of the area providing two complementing perspectives. First, from the end of the process, to understand the current usage of journalistic maps, which also informs on trends and possible unexplored design choices. Second, from the beginning of the process, to understand the day to day practice and difficulties in the production of journalistic maps. To answer **RQ1** we comprehensively collected and analyzed a large corpus of 462 journalistic maps from five major newspapers published over a period of three months. As a result of this analysis, we created a design space comprised of eight dimensions that involved both properties describing aspects of the articles and also the visual and interactive features of the maps in our corpus. We approach **RQ2** via a series of semi-structured interviews with four data journalists from a different range of media outlets. Through these interviews, we were able to identify the most common aspects and potential gaps in current practices. We also collected the practitioner's feedback on our design space from those interviews, which resulted in refinements to the design space, as well as comments on how it could be part of the journalistic map design process and even on the education of new data journalists. Our main contributions are:

- An empirically derived design space of journalistic maps;
- A set of insights on map production in the context of news;
- A set of artifacts documenting our analysis including the maps collected, a coding spreadsheet, and a codebook.

We hope that these resources will support practitioners and researchers in the design and research of journalistic maps.

## 2 RELATED WORK

Given the multidisciplinarity of our work, we review research on the use of maps for storytelling and journalism. We also position our research in the context of other efforts concerned with design spaces and surveys on the use of maps in news articles.

- *Arlindo Gomes is with Universidade Federal de Pernambuco.*
  *E-mail: agsn@cin.ufpe.br*
- *Emilly Brito is with Universidade Federal de Pernambuco.*
  *E-mail: ecab@cin.ufpe.br*
- *Luiz A. Morais is with Universidade Federal de Pernambuco.*
  *E-mail: gusto@cin.ufpe.br*
- *Nivan Ferreira is with Universidade Federal de Pernambuco.*
  *E-mail: nivan@cin.ufpe.br*

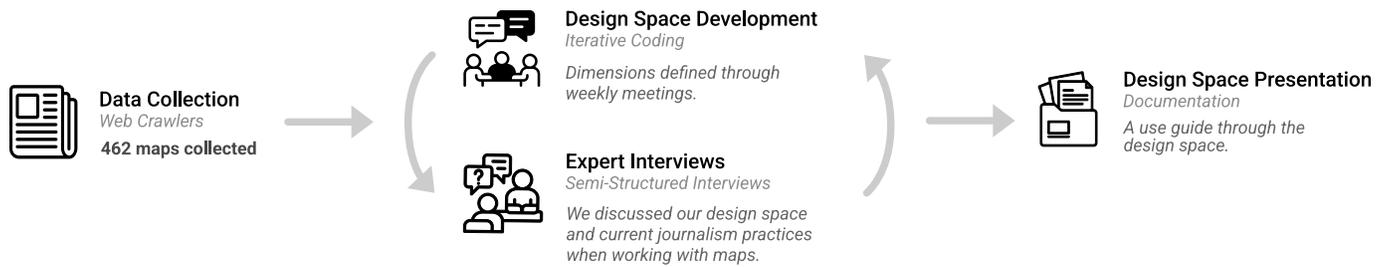

Fig. 1: Methodology of our study. In the first phase (Data Collection) we have compiled a corpus comprised of 462 maps from articles published in major news outlets. Afterwards we derived a design space via iterative coding sessions with the authors (Design Space Development) and semi-structured interviews with professional data journalists (Expert Interviews) . These interviews also included discussions on the experts' daily routines as data journalists, focusing on the map design process. Finally, we constructed the documentation of our design space that includes the dataset of collected maps, a codebook, and our coding spreadsheet. Icons provided by Noun Project Artists: Rena Putri, Nawicon, Yoyon Pujiyono, and Rusmicon.

### 2.1 Maps for Storytelling

Maps are powerful means to construct narrative visualizations [21, 38, 48]. In fact, they have been used for this purpose for a long time, from the seminal maps by Charles Minard [19] to modern digital maps used nowadays on news outlets to convey a plethora of subjects, ranging from election pools, military conflicts, disease outbreaks, and others. Story maps are any cartographic representation that presents narrative elements. They are often part of a visual story that can also include illustrations, graphics, photos, and videos to paint a larger picture of the events described in the narrative [43]. These maps can help frame the narrative by linking it to a location, to increase the realism of the story, and also stimulate narratives [12]. All of this makes story maps very versatile, being used in many narrative genres, including fiction, nonfiction, news stories, and even in education [28, 32]. Many studies focused on the usage of maps for storytelling. For example, Roth [44] proposed a taxonomy of genres for storytelling with digital maps. They leveraged the genres proposed by Segel and Heer [48]. Song et al. [50] proposed an empirical approach to investigate how map-based stories are designed and perceived by audiences. In particular, they investigate how aspects like narrative theme, features of the audience, visual storytelling genres and tropes influenced the users' overall retention, comprehension, and reaction. One important design consideration in the design of digital story maps is the use of dynamic elements. For example, Monmonier [39] highlights how data-driven animation could be a game-changer approach when associated with visual stories. Amini et al. [3] performed a questionnaire-based study to assess how animation techniques can affect viewer engagement.

While many advances have been made, the multitude of factors that influence their effectiveness and broad applicability make the understanding of design and usage of journalistic maps an open problem. This work aims to contribute to this context through a reflection on the current usages of maps and the current practices for their design in the context of data journalism.

### 2.2 Geojournalism: Maps in Journalism

Data journalism is an approach that is growing among newsrooms of all sizes. Journalists can integrate the data to tell complex stories and present factual data that corroborates the story being told [9, 14]. One recurring activity in the context of data journalism is the creation of maps, which are often featured in the news. Given the advances and availability of map-producing technologies and also the importance of maps for journalism, the design and implementation of these maps have been given the status of a subfield of data journalism called Geojournalism [16]. This area combines geographical data with interactive visualization and storytelling techniques to enhance the spatial understanding of the story, improve engagement, and guide readers' perception of complex themes. For example, during the COVID-19 pandemic, maps were a standard tool for journalists to report on the spread of the disease, locally and globally [34]. Maps were a natural choice as they can be visually appealing while maintaining the informative nature of the news [10].

The practice of geojournalism is challenging due to many aspects related to the journalistic environment and application context. First, given that news is commonly intended for the general public, one of the main concerns is related to data and visualization literacy [45]. Those refer to the capacity to read, work, and analyze data critically, a skill that enables the readers to transform data into knowledge [22]. Furthermore, journalists are often faced with tight deadlines and high pressures to complete their work. Finally, the challenge of building data-driven stories is a highly multidisciplinary issue. Data journalists build not only stories, but are faced with the design of visually and spatially rich experiences [46, 51, 57]. All of these result in little diversity in the applied maps, with the interactivity mechanisms and visual metaphors often following previously established templates [4, 36]. This issue has long been a concern for researchers in this field. In fact, 40 years ago, Gilmartin [21] already questioned the limits of how much a journalistic map should be facilitated. How much does it need to appeal visually? What are the guidelines for constructing maps? Those problems remain challenging to this day, as many factors can impact the final reader's experience. Obtaining users' feedback is a meaningful way to assess the effectiveness of particular designs and can highly impact journalistic production [29].

### 2.3 Design Spaces of Maps in News Articles

Some works have focused on understanding the design of data-driven news articles, which include the use of visualization elements. A worth mentioning example is the work of Hao et al. [23]. They used news articles on the COVID-19 pandemic to derive a set of design patterns that include both narrative and visualization aspects. However, while they discuss general challenges for data journalists and visualization designers and also mention maps as one possible visualization in news articles, the design of these maps is not detailed. Zhang et al. [58] derived a design space for geo-infographics. For this, they surveyed 118 sample infographics and also eight sketches by experts. They also introduced an authoring tool based on the design space with the goal of better supporting the design of infographics, even for users with little experience in infographics design. While many infographics appear in news articles, our work differs in that the goal was not to put the map design and data journalist experience into perspective. Mayer et al. [33] propose a design space of spatiotemporal visual data stories. To do this, they analyzed 130 web-based data stories with spatiotemporal characteristics, all gathered from news outlets in a period of two years. Since the focus is on the spatiotemporal aspect of the visualization, again, the focus is not on maps; for this reason, many aspects concerning map design are not present.

Closer to ours is the work by Pretsby [43]. He used different taxonomies to classify 117 journalistic maps that are present in news about the COVID-19 pandemic. This classification focused on aspects such as story context, genre, trope, trope techniques, and vividness [18]. In particular, he used the genre and tropes for maps defined by Roth [44] to classify the journalistic maps. The goal of his work was to identify common design traits in the news stories presented. Our work has a different goal: to understand in detail the design of maps used in news articles and to provide a journalist's perspective on this task. Furthermore, unlike the works mentioned, we don't focus on a particular theme; instead, we compiled a comprehensive corpus of map usage for a set of media outlets and a specific period.

## 3 METHODOLOGY

We now describe the methodology of our work, which is composed of four stages. In the first one (*Data collection*), we have compiled a corpus of maps published in major media outlets. We then analyzed this corpus (*Design Space Development*) to derive a design space that describes current patterns in the design of journalistic maps. This analysis was coupled with semi-structured interviews (*Expert Interviews*) with data journalists about their practice and also served as a way to refine our design space iteratively. Finally, we compiled our findings (*Design Space Presentation*) as the final version of the design space and additional materials to support further research in the area.

### 3.1 Data collection

The primary focus of our work is to understand how journalists utilize maps as supporting tools for the main story of their articles. For this reason, our main goal in this stage was to collect journalistic maps that were representative of the map design used in media outlets, with a focus on newspapers. Mindful that this is a demanding task given the wide variety of media outlets throughout the world, we tried to mitigate possible biases in two ways. First, we considered relevant outlets that are widely recognized and read globally. These outlets are often the ones that set trends in article visual design. Second, we did not restrict our collected articles (and maps) to a particular subject, but instead tried to perform a comprehensive search on all the published articles.

The choice of which media outlets to consider was guided by the work of Ausserhofer et al. [5], which, as a result of an extensive literature review, lists the news outlets that were considered in works in the area of data journalism. We selected the top 10 cited in this work. However, we eliminated five of them due to incompatibilities with the web crawlers, not being in English, or not functioning outside of the USA. The remaining five were included and are: Financial Times, New York Times, Washington Post, The Guardian, and ProPublica.

To form the initial list of news articles, we utilized the Google News Scraper (GNS) package [15], which enables the automatic search and saving of Google News results. We employed operators to constrain the search to specific websites and time intervals. While GNS worked well in all the outlets, it faced restrictions on The New York Times, which restricted the use of search operators from Google and crawlers in general. However, they provide the New York Times Article Search API [52], which supports the same functionalities as GNS and was used in this work. Despite the approach using GNS and NYT Search API being successful in gathering articles with maps, post-scraping was necessary to clean the false positives, as the list of news articles contained pages that did not include maps or pages that were not news articles, like front pages and pages that aggregate articles from the week. Our strategy to find journalistic articles that contained maps was to simply search for articles that included the word *map* in them. This choice was made since, in all our observations, the maps are explicitly referenced in the text as "map". Furthermore, we set the timeframe for our search from February 1st, 2024, to April 30th, 2024. This period was chosen to allow both a reasonably large number of maps and also to provide enough time for the execution of the rest of the study. As a result of this initial search, we obtained 1831 articles. To build our corpus, we manually inspected all the articles and gathered the maps on them. We identified numerous instances of repeated maps (multiple articles reusing the same figures) and false positives (articles containing the word 'map' without accompanying map visualizations), which were subsequently removed. Furthermore, we defined the following inclusion and exclusion criteria for the maps we found:

- *The article must not be about the map itself:* Articles where the maps are the main point of discussion are not what we are looking for. Our focus is on understanding how the map supports the article to tell stories. Therefore, our goal was to exclude articles that did not focus on telling a story. For example, articles that critique specific maps or their colors, quality, projection methods, and other aspects were removed.
- *The maps should represent real places and real data:* We want to ensure that our corpus reflects geographically accurate data and is tied to actual events. Therefore, we removed maps that depict fictional worlds from movies, TV shows, cartoons, or simulations with no specific geographic localization.
- *The maps should be published to inform about a subject:* The maps gathered should serve a clear journalistic purpose, such as explaining conflicts, health hazards, arts, lifestyle, and others. With this criterion, we aim to exclude maps that do not support stories, such as those built from user engagement (e.g., Snapchat, Instagram maps, or Yelp), maps used in advertising images, or educational purposes (e.g., quizzes or games).

After applying these criteria, we identified 358 articles and 462 maps. We highlight that this number is considerably higher than the ones considered in previous works that studied data journalism [23,33,41,43]. However, we acknowledge that the results may still be biased by the events that occurred during this timeframe and the editorial choices of the news outlets we collected. We manually collected the maps and the accompanying data representations as images, highlighting the presence of interactivity and media to make this information easily accessible during the coding phase.

We developed a code to reference the maps we found, which is based on three elements: the news outlet, the article ID (relative to that outlet), and the map ID (relative to that article). For example, the code **[FT001-02]** refers to a map found in Financial Times, in an article which was attributed the ID *001* and is the second map in that particular article *02*. The last portion of the code can be discarded when there is only one map in the article. In this case, we use the code **[FT001]** instead of **[FT001-01]**. The list of the collected maps alongside their codes is provided as supplemental material. We will use these codes to refer to particular maps in our corpus through the paper.

### 3.2 Design Space Development

To develop our design space, we performed an extensive literature review considering previous design spaces in data journalism, cartography, and map design (see Section 2). Furthermore, we used the previous experiences of the authors—designers, journalists, and visualization experts—to derive an initial set of dimensions.

The coding process began with a rater calibration, during which the first two authors independently coded and compared 30 items from the corpus to align their approach. After this, they divided the remaining maps by news outlet, and each coder classified a separate subset. The first two authors then reviewed each other's classifications, followed by a final review of all coding results by the remaining two authors. The dimensions and sub-dimensions of the design space were iteratively refined during meetings with the first two authors and in weekly meetings involving the whole team. During these meetings, the team evaluated the dimensions, updating them to reflect the impressions of the interviewed journalists and addressing potential coding conflicts identified by the authors.

The previous phases resulted in refinement of our related work and in a series of artifacts that compose the **design space presentation**. The supplemental material includes a codebook that provides assistance in understanding how to classify the maps into the dimensions, our coding spreadsheet listing all the maps and their article links for visualization and further research, and finally, a document containing images of all the maps collected in our corpus. Section 4 describes the final version Design Space.

### 3.3 Expert Interviews

To understand the daily practices of map design in data journalism and the feasibility of our design space in newsrooms, we conducted semi-structured interviews [1] with four Brazilian data journalists who create data-driven news articles daily. P1 works for a small newsroom specializing in news and reports about the Amazon region. P2 creates maps, infographics, and illustrations for a medium-sized independent news outlet that primarily produces data-driven articles, 30% to 40% of which contain maps. P3 belongs to a data visualization team at one of Brazil's most prominent news outlets. P4 works as a freelancer designing maps and infographics for various news outlets and non-profit organizations. The interviewees work on one to three articles per week, according to the complexity of the visualization and story. After that, we performed a Thematic Analysis [13] to build a structure that can help us understand the main hurdles and challenges involved in designing journalistic maps.

The meetings were held online for 60 to 90 minutes and conducted in three phases. First, we made three open-ended questions about their educational and professional experiences, the dynamics of production, and the tools they commonly use to build maps. In the second one, we did a presentation about the design space and presented dimension by dimension, showing and collecting feedback about it. Finally, in the third phase, we asked them if they noticed how to apply our design space in their professional context.

We then transcribed and performed a thematic analysis on the interviews (see Section 6). Using this approach we were able to determine the main themes discussed during the meetings. Furthermore, their feedback helped us refine our design space and understand its place within the newsrooms.

## 4 DESIGN SPACE OF JOURNALISTIC MAPS

We conceived a design space of journalistic maps for news storytelling based on the characterization of 462 maps from five major news outlets that work with data journalism, shown in Figure 2. This section contributes to answering the research question of *"What is the design space of journalistic maps?"*. Our design space contains eight dimensions corresponding to map and article features. Some dimensions were inspired by the works of Hao et al. [23], Munzner [42], and Mericskay [35].

### 4.1 Map Features

Creating modern data maps requires design decisions regarding different levels of information and interaction. Designers need to abstract the topic and dataset into the map's visual representation, consisting of its physical and thematic layers [25, 37], to establish how the map fits into a story and which data attributes are important. The *physical layer* represents geographical and topographical features, such as political boundaries, vegetation, and roads, forming the **base map** and providing spatial context. The *thematic layer* contains data that conveys specific information based on the map's purpose, such as population density, climate zones, or electoral votes. To build the thematic layer, a designer considers decisions regarding **map type** and **information overlays**, which help users extract insights from spatial data. Finally, we identified that computer-supported maps can also have an *interaction layer* that enables users to dynamically alter the physical and thematic layers via **interactive actions**. In the following subsections, we provide an analysis of the map features in our corpus.

#### 4.1.1 Base Map

Base maps serve as a reference map, consisting of the physical layer of a particular location. We identified base maps with varying levels of detail. **outline maps** (N=274|59.3%) are solely composed of lines corresponding to geographical boundaries like countries, regions, or cities. **tilemaps** (N=98|21.2%) are a concatenation of small area squares (i.e., tiles) representing detailed geographical information about the built environment, terrain, and water bodies. **relief maps** (N=119|25.8%) show the variations in slope, aspect, and elevation of a specific area, representing hills, valleys, and mountains. **satellite maps** (N=24|5.2%) provide a detailed visual representation of a region, built from a compilation of high-resolution satellite images. Lastly, **3D maps** (N=17|3.7%) represent a group of maps that include 3D relief maps, 3D photogrammetry, 3D photorealistic maps, and 3D tilemaps. In our analysis, we also found one example of a map that does not contain a physical layer, coded as [FT056], which we call a **blank map**.

#### 4.1.2 Data Dimensionality

A map that displays only raw geographical information (e.g., political borders) and no thematic data is known as a *reference map*. In contrast, *thematic maps* also visualize thematic attributes, which are spatial variables connected to locations on the base map [25, 37]. Our work identified different levels of data dimensionality in thematic maps, which means the number of data attributes chosen by the designer to tell a story. **univariate maps** (N=156|33.8%) represent a single data attribute, such as the temperature in a weather forecast map. **bivariate maps** (N=45|9.7%) encode two data attributes at the same time, such as rain vs. crop quality in a region. Finally, **multivariate maps** (N=38|8.2%) can represent more than two data attributes as in [FT014].

#### 4.1.3 Map Type

The map type refers to the methods used to represent thematic data and communicate spatial relationships, patterns, and trends. A **dot density map** (N=28|6.1%) combines point marks to visualize the geographic distribution of a particular data attribute. **heatmaps** (N=64|13.9%) and **choropleth maps** (N=54|11.7%) are both used to represent spatial data with color. The main difference is that the former emphasizes data density without boundaries, while the latter highlights differences across defined regions. There are also map types that represent time-related changes. **temporal maps** (N=48|10.4%) are designed to show how a phenomenon evolves across a geographic area. They can represent changes such as population shifts, climate patterns, or urban development by displaying sequential snapshots as [FT121-01] or continuous animations like [FT089]. **flow maps** (N=49|10.6%) represent the movement of people, goods, or information between different locations using directional lines or arrows. Although they can incorporate a time element to show how the flow changes over time, their primary purpose is to represent directional connectivity and the magnitude of the flow. One of the simplest map types is the **locator map** (N=203|43.9%), which highlights specific political or geographic regions to draw the viewer's attention towards them. Locator maps can be considered reference maps since they do not contain a thematic layer. The map types that did not correspond to at least 5% of the corpus were excluded from the final representation of the design space; these included juxtaposed maps, area maps, roadmaps, illustrations, numeric maps, scaled cartograms, proportional symbols, and network maps.

#### 4.1.4 Information Overlays

Overlays are components displayed on the map to bring additional information or clarification about specific physical or thematic attributes. Minimaps and inset maps are overlays that offer alternative views of a particular region. While **minimaps** (N=175|37.9%) show the overview of an entire area to help users navigate a comprehensive space without losing the focus on a specific location as in [WP035], **inset maps** (N=16|3.5%) use zoomed-in details (e.g., a city's downtown within a state map) or zoomed-out views (e.g., a country map showing its location within a continent) to provide additional context about specific parts of the map as in [FT001-01]. Other overlay types include further information on particular parts of the map to enrich the map's narrative. **geographical labels** (N=368|79.7%) (e.g., city names, landmarks, etc.) help the user situate the story to a particular place. Likewise, some designers use **icons** (N=19|5.3%) to represent relevant data linked to specific locations on the map. The designer can also utilize **map annotations** (N=134|29.0%) to clarify specific events or include **embedded images** (N=7|1.5%) on the map to illustrate these events as in [WP014]. Finally, a data-driven map usually has a **legend** (N=123|26.6%) to explain the meaning of visual encodings.

#### 4.1.5 Interactivity

An advantage of web-based news articles is their ability to engage readers with the data through interaction. We identified interactivity actions that aim to enhance the storytelling experience. Journalists often try to provide facilitated access to the stories. One common method is scrollytelling, where users can *navigate* the data stories with **scrolling** techniques (N=30|6.5%), where the page smoothly changes and

## Data Dimensionality

The number of variables that the analyzed map represents, this choice can directly affect the complexity and readability of the map.

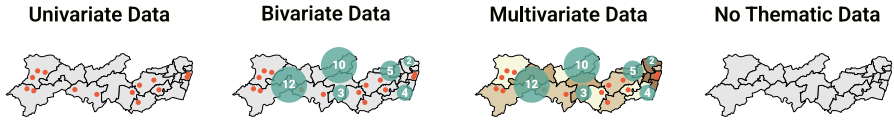

Univariate Data | Bivariate Data | Multivariate Data | No Thematic Data

## Interactivity

Interactions applied to the web-based maps, resources that allows the reader to explore, manipulate, reveal and filter the data showed in the maps. They are separated in Three groups:

*Explore*

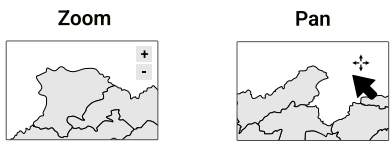

Zoom | Pan

*Navigate*

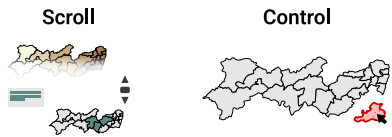

Scroll | Control

*Focus*

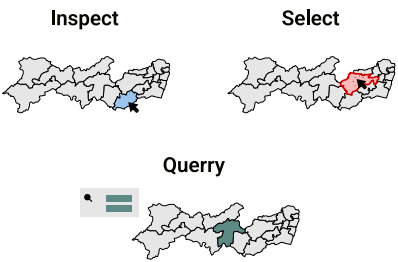

Inspect | Select

Querry

## Article Type

The main objective of this dimention is to understand the analytical depth applied to the news article and the main narrative objectives that drives the news article.

**Quick Updates** - Articles that are a snapshot of the story or the data, often used as a quick format to present daily and urgent news.

**Briefings** - Articles that present summarizations and overviews of events in a specific timeframe.

**Chart Description** - Articles that present a direct narrative approach, were data is used to do comparisons and to support the narrative that is being constructed through interconnected visualizations.

**Investigation** - Articles that are focused on complex storytelling. They try to expose correlations, trends and different viewpoints, and often present interviews associated to the narrative.

**In-Depth Investigation** - Also focused in presenting complex narratives but with a more comprehensive data analysis, involving multiple datasets, various types of visualizations and a stronger statistical analysis.

**Article Features** | **Map Features**

## Information Overlays

Overlays are information displayed on top of the map; they provide context and help to situate the readers on the map, helping to build a complete visual narrative.

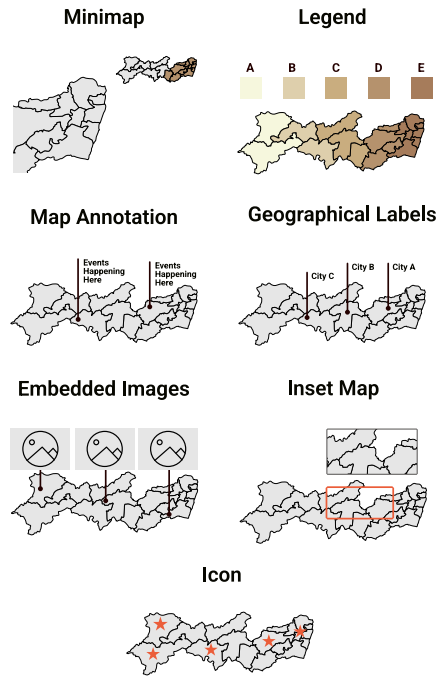

Minimap | Legend

Map Annotation | Geographical Labels

Embedded Images | Inset Map

Icon

## Acc. Data Representations

Those are data plots that can be added to work alongside the maps building the news article narrative.

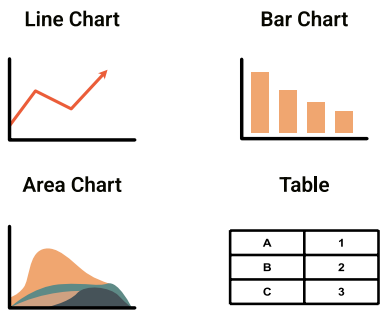

Line Chart | Bar Chart

Area Chart | Table

**Other**

*Timeline, Proportional Area Charts, Lollipop Chart, Bubble Chart, Scatter Plot, Slope Chart, Tree Maps.*

## Base Map

The fundamental cartographic layer where the data is overlayed defines the spatial context and representation of the map

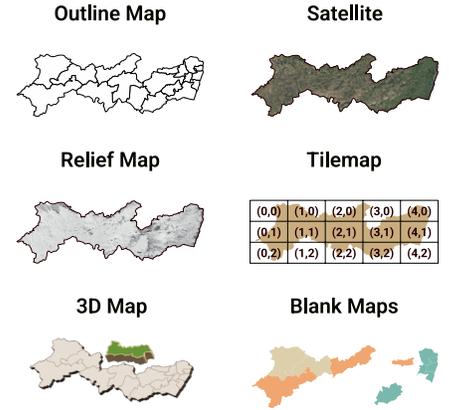

Outline Map | Satellite

Relief Map | Tilemap

3D Map | Blank Maps

## Map Type

The chosen cartographic method to represent the narrative, considering its purpose and visual design. Each type of map communicates different aspects of the news and can support different lenses in the narrative.

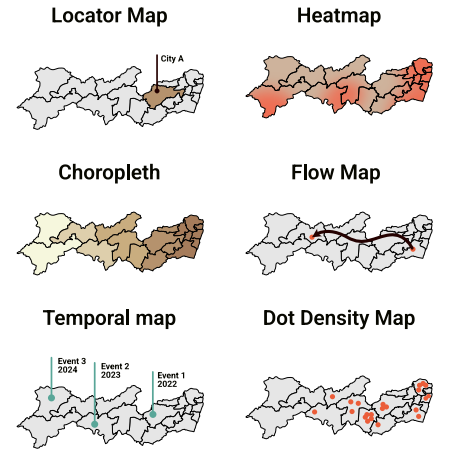

Locator Map | Heatmap

Choropleth | Flow Map

Temporal map | Dot Density Map

**Others**

*Juxtaposed maps, Area Map, Roadmap, Illustration, Numeric map, Scaled Cartogram, Proportional Symbol and Network Map.*

## Supporting Media

Representative media that is inserted in the news articles, serving as media assets that provide depth and visual support to the articles.

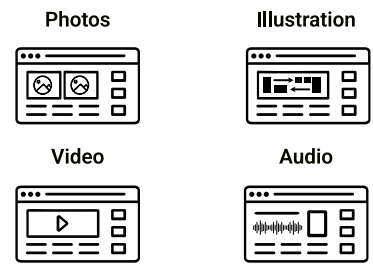

Photos | Illustration

Video | Audio

Fig. 2: Design space of journalistic maps in news storytelling.

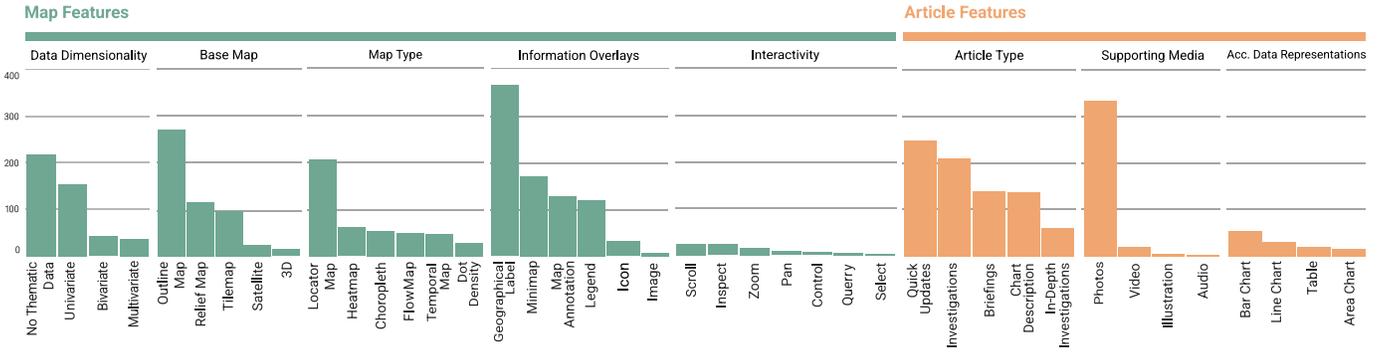

Fig. 3: Distribution of design space dimensions across the 462 maps in our corpus.

the narrative unfolds in a controlled rhythm as we can see in [NY006-01]. Readers can also use input elements such as sliders or buttons to **control** (N=6|1.3%) the flow. Regarding map interactions, readers oftentimes *explore* an area by **zooming** (N=16|3.5%) and **panning** (N=9|1.9%) it. Finally, people can *focus* on a specific map content using different interactivity actions. They can choose to analyze a data subset by **selecting** (N=3|0.6%) a given area on the map or focus on a specific target and **querying** (N=4|0.9%) a value in a list, drop-down, or textual search. Finally, people can also **inspect** (N=24|5.2%) particular information on the map by hovering over labels or icons.

## 4.2 Article Features

News articles follow certain design principles to establish a narrative and a goal. A designer usually chooses an **article type** to decide how the story will be framed. We identified that, besides the map, additional elements could be included in the article to support map-driven stories, namely **accompanying data representations** and **supporting media**. In the following subsections, we describe the article features and the number of articles in which they occur.

### 4.2.1 Article Type

The article type refers to the rationale when constructing a news article containing maps and its narrative goals. This dimension is directly influenced by Hao et al.'s work [23]. **Quick update** (N=119|33.1%) articles provide minimal text and include maps that represent a snapshot of current data regarding daily or urgent topics such as weather events or politics. **Briefings** (N=38|10.6%) present maps that give an overview of events within a specific time frame without delving into detailed interpretations. **Chart description** (N=55|15.3%) articles use multiple maps as a medium to support a data-driven narrative, using them as support for comparisons and specific topics. Finally, **investigation** (N=104|29.0%) and **in-depth investigation** (N=18|5.0%) articles involve the use of maps and other accompanying visualizations to support complex storytelling. While investigation articles have a qualitative approach, focusing on interviews and the use of multimedia to tell stories as in [FT004], in-depth investigation articles use multiple data sources and a diverse set of maps and visualizations to support the narrative, as shown in [WP021-01].

### 4.2.2 Accompanying Data Representations

Some articles incorporate data representations along with maps to enrich the narrative and data analysis. We identified in our corpus charts that function as accompanying data representations such as **bar charts** (N=47|13.1%), **line charts** (N=30|8.4%), and **area charts** (N=9|2.5%). Additionally, it is common to use **tables** (N=11|3.1%) to display information. We also found the occurrence of other visualizations, but they only count for 5% of the corpus. They are timelines, proportional area charts, lollipop charts, bubble charts, scatter plots, slope charts, and tree maps.

### 4.2.3 Supporting Media

It is common for newspapers to provide additional context to news stories through representative media. We found that articles usually combine maps with **photographs** (N=342|95.3%) or **videos** (N=19|5.3%). We also identified **illustrations** (e.g., infographics, comics, drawings, etc.) in four articles and **audios** in two, [NY0133] and [WP031], which seems to be a less common supporting media.

## 5 Corpus Overview and Design Patterns

This section examines the characteristics of our corpus, focusing on common design patterns in the physical, thematic, and interaction layers of journalistic maps and their relationship with article features. See Figure 3 for more information about the occurrence of maps in each dimension.

### 5.1 Maps Across News Outlets

The five news outlets analyzed use maps in their articles at different frequencies. The Financial Times stands out with the highest volume, publishing 190 articles featuring maps and a total of 211 maps over a three-month period. The Guardian frequently incorporates maps, with 90 articles and 92 maps, averaging nearly one article containing a map every day. The Washington Post also heavily relies on maps, showing 77 maps in 60 articles. The New York Times has the highest maps-to-article ratio, with 89 maps appearing in just 52 articles, meaning they tend to include multiple maps within a single article. Conversely, ProPublica has the smallest sample, with three maps across three articles. These findings indicate that while major news organizations frequently integrate maps into their narratives, independent media may employ them more sparingly.

### 5.2 Frequent Topics

Our corpus covers a wide range of topics. The prominence of maps in geopolitics (N=173|48.2%), economics (N=39|10.9%), environment(N=33|9.2%), weather (N=39|10.9%), tourism (N=33|9.2%), and elections (N=31|8.6%) likely reflects both global events from early 2024 and the editorial priorities of the selected news outlets. The prevalence of geopolitical and election-related coverage aligns with ongoing international conflicts (e.g., the Ukraine War) and the U.S. presidential primaries. The presence of articles related to tourism in the corpus may signal post-pandemic industry recovery and rising interest in eco-tourism, for example. The inclusion of data from the Financial Times likely skewed the corpus toward economics-related articles. Finally, the weather-related articles in the corpus, particularly on climate change, reflect The Washington Post and The Guardian's environmental focus.

### 5.3 Supporting Media and Accompanying Data Representation in Different Article types

The majority of the articles widely use photos to add depth to the news as complementary visual context, with (N=433|93.7%) articles presenting photos. However, videos and illustrations are less common in the articles. Those kinds of supporting media were more common in quick updates (N=10|2.2%) and investigation (N=16|3.5%) articles. In

most cases, the photos have only a contextual connection with the maps, both supporting the same narrative. However, there are cases, such as [FT096] and [TG011], where they directly display the same event or highlight the same structure, respectively.

Investigation articles were the ones that most included accompanying data representations (N=40|8.7%), followed by chart description articles (N=26|5.6%).

Unlike in supporting media, it was more common to see cases where journalists connected the map and accompanying data representation, showing the same data with different representation approaches. Cases like [WP010-01], where the data shown on the map is detailed in the visualization in the article. In the case of [WP021-01], the data on the choropleth map is organized by state and is expanded in the bar charts into different demographic aspects (such as gender, ethnicity, and age) using the same colors as the map. In [FT083], we notice a different approach. While the maps showed values across countries, the bar + line timeline visualization focused on a specific city, and a treemap concentrated on the countries.

### 5.4 Common Map Design Patterns in News Articles

Data journalists structure articles to align with their message and goals [23]. Therefore, map usage patterns can vary according to subject, news outlet, or editorial decisions. As a general characteristic of the design patterns from our dataset, we can say that most articles in our corpus showcase locator maps (N=203|43.9%). Most maps in our corpus are static (N=397|85.9%). However, when interactivity is present, we identified that patterns could be tied to particular map types. For example, locator and dot density maps often include interfaces to help users navigate the map, typically employing scrollytelling (N=30|6.5%) and inspect (N=25|5.4%).

**Quick Updates** is the most frequent article type in our corpus (N=145|31.4%). In these articles, locator maps are the most common (N=82|17.7%). Furthermore, maps in quick updates usually feature geographical labels (N=115|24.9%) (e.g., city names), and minimaps (N=63|13.6%), or icons (N=13|2.8%) to situate the story to a particular place. This kind of map may also have annotations (N=34|7.4%) to provide additional context to the story and legends (N=27|5.8%) to clarify the data, as seen on [FT196]. In this article type, journalists utilize base maps, including outline maps (N=90|19.5%), relief maps (N=37|8.0%), and tile maps (N=28|6.1%). Most maps in this article type are static (N=135|29.2%).

**Investigation** articles are the second most common article type (N=127|27.5%) and present similar patterns to quick updates. Again, the most common map type is locator maps (N=76|16.5%). Inherently featuring geographical labels (N=110|23.8%), minimaps (N=41|8.9%), or map annotation (N=28|6.1%). They usually utilize outline (N=85|18.4%) and relief maps (N=51|11.0%). Most of their maps are static (N=107|23.2%), but we also see occurrences of scrollytelling (N=12|2.6%) as [WP001] and inspect (N=7|1.5%) interactions like in [FT100].

**Chart descriptions** are very illustrative articles and presenting (N=77|16.7%) maps. Those maps usually represented univariate data (N=33|7.1%). They are the article type with the higher map-to-article ratio, featuring 1.4 maps per article. Those articles present heatmaps (N=20|4.3%), temporal (N=13|2.8%) and choropleth (N=12|2.6%) maps, even combining them as in [FT002-01]. Additionally, these maps frequently include legends (N=40|8.7%) to aid interpretation. As base maps, they often employ outline maps (N=50|10.8%) or tile maps (N=70|15.2%). Most maps in the chart description are static (N=72|15.6%).

**Briefings** presented (N=74|16.0%) maps, the majority of them representing univariate (N=33|7.1%) or bivariate (N=14|3.0%) data. On this map type, we notice a greater presence of flow (N=19|4.1%) and temporal (N=16|3.5%) maps, as briefings often present spatiotemporal data, as we can see on [NY019-01]. Frequently, these two map types rely on annotations (N=35|7.6%) to assist readers in understanding how events develop over time, as illustrated in [FT005-02]. Journalists often employ line-based maps in briefings (i.e., outline map (N=28|6.1%) or tilemap (N=24|5.2%)). Interactivity is applied usually through zoom (N=8|1.7%) and scrollytelling (N=7|1.5%).

**In-depth investigation** is the smallest article type in the corpus (N=30|6.5%). Considering data dimensionality, they usually present univariate (N=11|2.4%), no thematic data (N=9|1.9%) or multivariate (N=6|1.3%) maps. The most common map types were locator (N=10|2.2%) and choropleth maps (N=7|1.5%), presenting geographical labels (N=26|5.6%), map annotation (N=11|2.4%) or legend (N=10|2.2%). They present base maps such as outline (N=16|3.5%) and relief maps (N=9|1.9%). Despite representing a smaller set of maps, they present the highest variation in interactivity, supporting exploration, navigation, and focus.

## 6 Production Pipeline for Journalistic Map Articles

We now describe the findings from the interviews with four data journalists, aligned with the main themes we collected in our thematic analysis. This contributes to answering the question *"how do editorial teams produce journalistic map articles?"*. The following subsections detail the process of producing journalistic map articles in newsrooms.

### 6.1 Story Conception

The production of a data-driven article typically begins with a relevant story. Journalists and editors *"define key questions to shape the narrative and outline what can be addressed through data analysis"* (P1). While journalists often focus on trending topics, they sometimes find space *"for curiosities and less serious matters"* (P4) in their articles. At this point, data journalists and editors meet designers, developers, and data scientists to refine the article's scope and available data. The theme of *Time limitation* was depicted by all participants; this theme relates to the lack of time to deliver each article and to learn new techniques and methods. Urgency is a natural characteristic of journalism. However, data-driven news presents more challenges than regular news, as P3 noted *"Even the creation of a simple agenda for a data journalism article can be a slower process, depending on the data collection, format, and analysis, this initial setting can take some time"* (P3). While the involvement of a diverse team can vary based on editorial priorities and newsroom resources, these initial meetings guarantee alignment on project goals and expectations.

### 6.2 Data Gathering

Once the article's direction is established, data journalists collect, clean, and analyze relevant datasets to extract valuable insights. Ideally, a diverse group of professionals is available to tackle specific challenges that may arise. As P1 exemplifies: *"I work specifically on data visualization [...]. Typically, a person in the team with a statistics background is responsible for connecting the data between statistical analysis and our story."* (P1). However, gathering and processing the right data can be challenging. P1, for instance, argues that *"One thing is the technique you want to use to visualize, and the other is the data available to produce that visualization. So, aspects as data granularity can limit how it can be translated into a visualization."* (P1). Finally, when journalists cannot find relevant data, it deters the development of a journalistic map: *"It's not always possible to create a map because the data isn't geographical and doesn't allow it."* (P2).

### 6.3 Map Design and Development

The data is translated into visual elements during this phase, emphasizing clarity and narrative coherence. The design process usually involves a data journalist iterating on the narrative, visualization, and media in checkpoint meetings. Regarding media, P4 noted that he would consider adding 'Audio' in the media dimension of our design space, as he had produced audio to insert in news articles for previous jobs, and in our corpus, we found examples that made us include it as a subdimension for media.

For this step of the process, the central theme was *Map Creation*, which relates to the choices they made during production, the types of maps, elements used, and the trade-offs from these choices. Journalists also consider the feasibility of creating maps given a deadline: *"Sometimes, utilizing a map is more laborious, as it often takes more than one map to display all the information"* (P2). When they decide to use maps, the team carefully selects which data points to highlight, ensuring that unnecessary details do not overwhelm the audience, as *"including unnecessary [visual] elements may hinder the [data] interpretation"* (P1). Additionally, one of the challenges for data journalists is *"making the visualization accessible, [...] regardless of the educational background"* (P1). Those are aligned with the theme of *Data literacy*, where participants explain their concerns and the measures they take to ensure that their stories are accessible and engaging for a broad audience.

Map design can vary depending on the article's priority and newsroom preferences. An urgent article, for instance, would require a simple map, while articles with a longer timeframe would allow journalists to iterate over weeks to design more detailed maps. Journalists frequently use locator maps to provide geographic context because they are easy to create. In contrast, P1 argues that 3D maps, though useful in some urban-focused stories, are rare due to their complexity and the resources required to produce them. Despite priorities, every news outlet has editorial and art direction preferences regarding map types and base maps to construct their stories. P2 explains that their graphics team prefers thematic maps over reference maps, but they constantly create locator maps because the newsroom *"likes to use them to support the audience in identifying which place the news is talking about."* (P2). Those design choices seem to reflect the nature of a newsroom environment.

Regarding interactivity, journalists prefer static maps. All participants highlighted that it is hard to explore solutions using interactivity, as their daily activities are overwhelming. P4, for example, declared that they do not use interactivity frequently *"because an interactive map costs time and money. For instance, you must test it on multiple screen sizes."* (P4). Financial decisions also impact the adoption of interactive maps and form the theme of *Budget Limitations*. As stated by P2: *"We use more static maps because the automated tools that facilitate producing interactive maps are usually paid, and we prioritize using as few paid tools as possible"* (P2). Incompatible skills also deter journalists from creating interactive maps: *"My lack of training on interactivity limits me a lot, but other people focus on building interactive visualizations on my team."* (P3) Regardless, P4 argues that creating static maps may be justified because *"from my experience doing visualization, media consumption nowadays is mainly passive and instant"* (P4). The implications include prioritizing more effortless visualizations to avoid disengagement: *"If we present friction, high cognitive load, or make the reader think a lot, most readers will lose interest"* (P4).

All participants used QGIS and Adobe Illustrator as their primary tools to produce journalistic maps. Other tools like Flourish, Datawrapper, and Blender were also cited as eventually being used by the participants. P1 and P4 explained that they always try to opt for open-source tools, as tools like ArcGIS and Mapbox are prohibitive in their context. Nonetheless, the journalists reiterated that it is possible to have good results despite using open-source tools. P1 and P2 mentioned that their teams occasionally construct tailored solutions for specific visualizations using JavaScript, R, or Python. Still, those projects are rarer as they demand someone committed to coding and maintaining the visualizations. All interviewees stressed that tight schedules also limit the production of articles with more interactive maps. P2 pointed out artificial intelligence tools as an alternative to ease the creation of such maps. Also, journalistic teams constantly try to improve their workflow by creating templates and searching for new tools to enhance their delivery of news.

# 7 DISCUSSION

## 7.1 Challenges and Opportunities for Journalistic Maps

We identified a set of challenges that data journalists have to overcome in the newsroom: tight deadlines, news outlets' financial strain, limited skills, and the impact of social media on news consumption. This section will discuss the issues associated with using maps in newsrooms and ways to enhance the production pipeline for journalistic map articles.

### 7.1.1 Tight Deadlines, Limited Skills, and the Modernization of Newsrooms

There is a duality between the urgency of news production and the meticulous process of creating data-driven articles. In our interviews, we discovered that editorial teams often ponder when it is worth investing in maps to tell stories, as they face tight deadlines and, sometimes, a single map is not enough to support a complex story. This hurdle is also cited by Houtman et al. [4], who interviewed 17 news cartographers from US and UK news outlets, including employees from three of our sources (The New York Times, Financial Times, and The Washington Post). Another similarity is that for Houtman's participants, web development for map interactions is described as time-consuming and difficult [4], a problem that was present in all of our interviews. Creation tools like *Mapbox* and *Mapcreator.io* assist journalists in building maps quickly. However, this approach might lead to the standardization of map formats, restraining unique and innovative design decisions [23]. Journalists must also tackle demanding technical tasks like finding data sources or performing data scraping to produce articles with journalistic maps. The recent popularity of artificial intelligence (AI) tools has helped data journalists create code for visualization, scrapers, or even new custom tools for the newsroom, aiding journalists to be more independent and act faster in the newsroom.

The widespread use of locator maps with no thematic data for quick updates and investigation articles is a common practice that dates back to the 80s, still corroborating with Monmonier [38], which explains that locator maps can play the role of a "cartographic icon" by supplementing the headline and setting the audience in the place. The journalists interviewed explained that the design choice sometimes comes as a request from other teams that use maps solely with this objective, but do not explore other possibilities within them. Another recurring pattern in news articles is the use of univariate maps. This trend relates to the complexity of displaying multiple variables together and the effort required to create accessible visualizations for multivariate data, a problem highlighted by Fairbarn et al. [17], where the authors noted that the use or combination of spatial data from different types, properties, and methods of data collection might require different representations, challenging the map to combine them, something that requires more time and effort than univariate small multiples for example.

Interactive visualizations are uncommon in smaller newsrooms, but even prominent news outlets—with dedicated teams to create data visualizations—are reluctant to deliver interactive maps. Scrollytelling ends up being the most adopted approach to provide interactivity. The wide use of this design pattern indicates that news outlets understand the need for interactivity in their articles but prioritize simpler interactions to ensure easy implementation and readability. Furthermore, the act of scrolling on social media and the trend of passive media consumption also reflect the format of current news articles. As Searles and Feezell [47] work highlights, more clicks and shares are a clear motivation to make news scrollable, leading editors to write headlines that attract more attention. Furthermore, complex interactions can be costly and require testing and maintenance across various platforms.

Finally, our corpus also lacks more examples of sophisticated base maps such as 3D models, We associate the technical skills to develop this type of map and the fact theis gap with the difficulties of having openly available 3D geometry for the location of the news, as it exists for 2D maps. y didn't are a good fit for every news. More importantly, Interactive 3D maps such as [WP028] are even rarer. Despite realistic 3D models being appointed to promote memory retention of information in educational contexts [49], Interactive 3D maps are not seen as suitable for general audiences, but adequate for experts [24]. We noticed through our interviews and corpus analysis that the decision of which map to use for each article type is not straightforward due to a myriad of aspects, including editorial decisions, team skills, type of news outlet, and budget. As a company that produces content, these organizations must leverage all these aspects to remain viable and competitive. Further studies should be conducted to investigate how to assist data journalists in creating more innovative maps concerning physical and thematic layers.

### 7.1.2 Data Literacy for General Audiences

Data journalists often show concern regarding the readability of their maps and visualizations. As such, data representations are designed for the general public, making it hard to create innovative maps without provoking a higher cognitive load or friction between the public and information. It was common to see design patterns aiming to improve the readability and accessibility of maps in our corpus. For example, temporal and flow maps usually present annotations to support the narrative. Those are efforts to guide users through maps and guarantee that the audience fully absorbs the stories.

According to our participants, none of their news organizations performed any assessment or research to understand the data literacy level of their audience. It would be crucial for each news outlet to have some profiling, as it is known from previous research that general audiences can have difficulty identifying and reading complex data visualization [8]. Our dataset might foster new research on data literacy using actual data from news outlets. It could be combined with data literacy assessment tools [20, 30] to investigate how news users read and what they understand about journalistic maps.

Educational curriculum deficits, generic visualization themes, and insufficient teacher training are causes discussed in the literature that could explain the current design patterns of journalistic maps. In the work of Lopes, Reznik, and Kosminsky [31], the authors propose a case study of situated visualizations as a final project for a course. practical approaches like that promote critical thinking over technical skills, guiding students to build their data literacy skills on themes they are acquainted with. Other approaches, such as Data Comics [6] could help teachers leverage visual storytelling and structured data narratives in the classroom. guiding the reader toward the story the visualization was made to tell.

## 7.2 Using the Design Space on the Newsroom

This section discloses how our work can positively impact newsrooms, independent data journalists, and map designers by providing a framework for conceptualizing and classifying journalistic maps.

### 7.2.1 Training Novice Journalists

Our design space can have a direct impact on novice data journalists' education; we view that as one of the key qualities of our work. The design space could be handy, as to the best of our knowledge, there are not many supporting frameworks for general map creation. This corroborates that most collected maps from our corpus were from locator maps with no thematic data—simple and quick maps to build, but often limited in depth and constructed with automated tools such as *Mapcreator*. The design space could also be used in classes to understand the elements of journalistic maps. It could support workshops and exercises in classes by helping students conceptualize or classify maps, generating debates about their elements, objectives, and characteristics.

### 7.2.2 Enhancing the Production Pipeline

For map designers in the newsrooms, one of the main benefits this design space could bring is related to the conceptualization of new maps, as it may enable designers to create maps and combine methods of map creation. Editorial teams could also use our work in team meetings to visualize and plan the visualizations that would be part of the article. Developing news articles is a collective process; more than one professional commonly works on an article. Using our design space as an enabler, we can build different plans for visualizations. For example, if we have bivariate data on transport in a specific street on a weekday and a Sunday, how can we show both information simultaneously? We could do side-by-side juxtaposed maps, an interactive map to select options, or an animated 3D map that switches between traffic jams and empty streets. Those design decisions could be discussed, considering the design space dimensions. This approach will require time and effort from the team, as planning the map alongside the visualizations must be done before the article is completed. According to our participants, this process generally occurs gradually, with one data journalist or designer typically leading this task. In the end, it becomes a trial-and-error process in most cases.

Additionally, our design space can also serve as a base for tools in map creation, including solutions that utilize AI, such as Aino [2], which can generate editable maps to provide designers with a starting map design. Regardless, our intention is not to make a rigid or automatic map creation method but to present a sandbox solution with possibilities to help designers build better maps based on their time, budget, data, and objectives.

## 7.3 Limitations and Future Work

Our first set of limitations is related to data collection methodology. Temporal bias may influence our findings, as certain events, crises, or social changes can disproportionately affect the coverage of specific topics, data types, and visualization techniques. Editorial bias is inherent in news organizations differing priorities and interests, shaping the subjects they cover. Locational bias arises from all analyzed news outlets being Western and publishing in English, which naturally skews the reporting toward their respective countries cultural and geographical contexts. Our reliance on Google News and The New York Times search results introduces potential algorithmic bias, as search rankings and filtering mechanisms influence retrieved results. Future research should address these limitations by expanding the scope to include news outlets from diverse linguistic, cultural, and geographical backgrounds, as these factors undoubtedly shape reporting practices and visualization approaches. While these constraints do not invalidate our findings, acknowledging them is essential to ensure transparency and guide future studies.

We chose not to perform overlapping between the coders, as we planned to perform dynamic code adjustments through group discussions. This limitation prevented us from using metrics as Inter-rater Agreement [27] and or Cohen's Kappa [56]. Additionally, our qualitative analysis is based on semi-structured interviews with four professionals specializing in data visualization. However, this sample does not encompass the full spectrum of roles within data journalism teams, such as data scientists, analysts, editors, developers, and crowdsourcing experts. Including these perspectives could offer deeper insights into collaborative workflows and the challenges associated with them. Furthermore, all interviewees work in Brazilian media organizations, where resources and team structures differ from those of international news outlets examined in our dataset. While many similarities were found with journalists from major sources [4], we acknowledge that expanding the study to incorporate data journalists aligned with the corpus or from other countries and contexts could reveal significant cultural and professional variations in journalistic data visualization practices.

## 8 CONCLUSION

In this paper, we describe how data journalists design maps to fit news articles and the decisions involved in building those maps in the newsroom. Based on an analysis of 462 maps from five major news outlets, we developed a design space encompassing the features related to article design—article type, accompanying visualization, and supporting media— and map design—data dimensionality, base map, map type, information overlays, and interactivity. This framework allows data journalists and researchers to conceptualize and evaluate map designs. To understand practices and decisions regarding the production of news articles with journalistic maps, we interviewed four data journalists from different backgrounds. Their feedback helped us to refine our design space dimensions and clarified how editorial teams create journalistic map articles. We hope that this material will help researchers and data journalists to use our design space in their context of work, and also provide them with a deeper understanding of map design for data journalism.

### ACKNOWLEDGMENTS

We would like to thank the reviewers for their constructive comments and feedback. We also would like to thank the data journalists that participated in the interviews. This study was supported by CNPq (311425/2023-2), and FACEPE (BIC-0682-1.03/24).